\documentclass[10pt]{elsart}
\usepackage[super, sort]{natbib}
\bibpunct{(}{)}{;}{a}{,}{,} 
\usepackage{amsmath,amssymb,subeqnarray}
\usepackage{graphicx}

\def\tableline{ \vskip .1in \hrule height.6pt \vskip 0.1in}

\begin{document}

\begin{frontmatter}

\title{A Method for Representing and Developing Process Models}
\author[cll]{Stuart R. Borrett\corauthref{cor1}},
\ead{sborrett@stanford.edu}
\author[cll]{Will Bridewell}
\author[cll]{Pat Langley}, and
\author[geo]{Kevin R. Arrigo}

\corauth[cor1]{corresponding author}
\address[cll]{Computational Learning Laboratory, CSLI, Stanford University, Stanford, CA, USA}
\address[geo]{Department of Geophysics, Stanford University, Stanford, CA, USA}

\begin{abstract}
Scientists investigate the dynamics of complex systems with quantitative models, employing them to synthesize knowledge, to explain observations, and to forecast future system behavior.  Complete specification of systems is impossible, so models must be simplified abstractions.  Thus, the art of modeling involves deciding which system elements to include and determining how they should be represented.  We view modeling as search through a space of candidate models that is guided by model objectives, theoretical knowledge, and empirical data.  In this contribution, we introduce a method for representing process-based models that facilitates the discovery of models that explain observed behavior.  This representation casts dynamic systems as interacting sets of processes that act on entities.  Using this approach, a modeler first encodes relevant ecological knowledge into a library of generic entities and processes, then instantiates these theoretical components, and finally assembles candidate models from these elements. We illustrate this methodology with a model of the Ross Sea ecosystem.
\end{abstract}

\end{frontmatter}
\newpage
\section{Introduction and Motivation}
Models are critical tools for determining how elements combine to generate the complex system dynamics that we observe in nature.  Ecologists use models to synthesize existing system knowledge into a concise form that guides empirical research programs \citep{osidele04, whip05}, but they also use statistical models to describe patterns in their data \citep{underwood97}, and simulation models both to explain and predict system behavior \citep{ford00, jorgensen01, melillo93, clark01}. Models also let scientists perform thought experiments that would not otherwise be possible or ethical.  Because of this advantage, ecologists have used models to build ecological theory \citep{abrams93, abrams00, jorgensen02, pulliam91, carpenter85} and to guide environmental assessment and management \citep{brando04, costanza98, jorgensen94, reckhow94, sage03, korfmacher01, maguire03}.

Quantitative models in ecology and environmental science are often categorized by the degree to which their structure corresponds to a real system \citep{levins66, levins93, orzack93, reckhow94, bossel92, hilborn97, zeigler74}.  This realism continuum begins with empirical models and ends with mechanisms.  Empirical models (\emph{e.g.}, regression-based models) stem solely from observed relationships among variables, provide a statistical summary of the data, and ignore mechanisms determining the behavior.  In contrast, mechanistic models contain unobserved relationships, explain system dynamics, and emphasize the physical, chemical, and biological processes that generate system behavior.  Ecologists use mechanistic models to understand how system behavior may change in response to changing environmental conditions.  A central problem of building models with more realistic structures is determining which entities and processes to include and which mathematical representation is most appropriate.

Following \citet{langley87}, we claim that model construction involves search through a space of possible models for ones that fit system observations.  This space contains alternative model structures (entities, processes and the connections among them), mathematical formulations, and parameter values. The immense number of possible models challenges scientists, who navigate this space by selecting the set of objects or entities used to represent the system and the relationships that link these objects to each other and their environment.  When establishing the entities, scientists must address three questions: (1) which entities should be included, (2) how detailed should they be, 3) and how should they be represented?  The answers determine object aggregations and system boundaries, both of which are rarely obvious and can significantly influence model results \citep{cale79,gardner82, rastetter92,loehle87, AA02, ahl96}.  After defining the objects, the scientist can state their relationships by deciding which ones to include and how to model them.  This task requires the specification of each relationship's mathematical formulation, which often involves selection from several possibilities.  For example, the Lotka--Volterra, Ivlev, or Holling Type III functions each model predation, but the different formulations make different claims about how the process operates.  Finally, the scientist must set the numeric parameters.  In some cases, empirical estimates of parameter values exist, but more often the values are largely unknown \citep{beck87,reckhow94}. Uncertainty enters the model at each decision point \citep{loehle87, jorgensen01, reynolds99, beck87, reckhow94}, complicating the search for plausible explanations.

Searching for models presents two additional challenges, one related to the selection criteria and the other centered on the search procedure.  Scientists require criteria for ranking and evaluating models \citep{oreskes98,oreskes_sci94,hilborn97,reynolds99,jost01}, which usually include one or more quantitative measures of goodness-of-fit of the predictions to observed data.  However, selecting a model based on its accuracy alone is insufficient since  very different models can generate similar behavior \citep{cale89}.  To overcome this problem, additional criteria such as a model's complexity, uncertainty and generality may be used.  If the model should explain system behavior, then its structure must also be sufficiently realistic \citep{levins66, zeigler74, bossel92}.  The second challenge is that the search procedure is cumbersome.  In ecology, this search is typically a manual effort guided by an expert's domain knowledge (\emph{e.g.}, aquatic ecosystems) and modeling experience, along with the data.   Given a model structure and mathematical formulation, some methodologies assist with fitting parameters and quantifying parameter uncertainty \citep[\emph{e.g.},][]{hornberger80, spear80, osidele04, saltelli00}.  Nevertheless, search through the space of model structures, mathematical formulations, and parameter values remains a challenging and time-consuming chore.

In this paper we describe a method for representing and building models designed to:
\begin{itemize}
\item facilitate construction of process-based models;
\item expedite search through the space of candidate model structures;
\item root model development in domain theory; and
\item bind models to empirical observations.
\end{itemize}
We first describe a new formalism for representing models as interacting sets of entities and processes.  We claim that this formalism captures how scientists understand complex system dynamics, and therefore eases model communication.  This representation also simplifies the comparison of a model's structure to the relevant domain theory.  We then show how this formalism facilitates search through the space of plausible model structures and illustrate the use of this approach by re-representing an existing model of the Ross Sea.  Finally, we discuss related work and propose some directions for future research.

\section{Process Modeling}\label{sec:frame}
In this section, we introduce the method for representing and constructing explanatory models.  The approach has two core elements: entities, which are the objects or actors in the system, and processes, which are the actions or activities of the entities that generate system dynamics.  Abstract forms of processes and entities encode domain knowledge, which is then used to construct models with realistic structure.  Scientists combine instantiated versions of the abstract elements to construct models of specific systems.  We conclude this section by briefly describing software we are developing to support this modeling approach.
\subsection{Entities}\label{sec:entities}
In process models, entities are actors and receivers of action that are characterized by a combination of variables and parameters.  For example, in a soil ecosystem model the collection of nematodes could be an entity with variables that describe its total carbon concentration and the number of individual organisms.  Depending on the model objectives and the processes included, a variety of parameters associated with the nematodes may be of interest including their maximum intrinsic growth rate, carrying capacity, and death rate.

In ecological models, entities are rarely differentiated from variables, which works well when there is only one state variable for each entity.  In such cases, ecologists commonly associate parameters with an entity by using the state variable's name as a subscript (see Section \ref{sec:RossSeaModel} for an example).  However, making entities explicit provides a natural way to group variables and parameters and more closely resembles how scientists think about real systems.
\begin{table}[t]
\caption{The syntax for defining a generic entity type.} \label{tab:ET-syn}
\begin{small}
\tableline{\obeylines\leftskip 2pt\parskip 0 pt %
entity \emph{name} \vskip 0.03in
~~description ``$description$''$^a$ \vskip 0.03in
~~variables \vskip 0.01in
~~~~~~$variable\_name_1$ \{\emph{combining scheme}\}$^b$, ``$description$''  \vskip 0.01in
~~~~~~$variable\_name_2$ \{\emph{combining scheme}\}, ``$description$'' \vskip 0.03in
~~parameters\vskip 0.01in
~~~~~~$parameter\_name_1$ $[range]^c$, ``$description$'' \vskip 0.01in
~~~~~~$parameter\_name_2$ $[range]$, ``$description$'' \vskip 0.03in
} \tableline
{\scriptsize$^a$Descriptions can be any text.\vskip 0.01in
$^b$Combining schemes state how the effects of multiple processes operating on a variables will be aggregated.\vskip 0.01in
$^c$Parameter ranges delineate legal values, which are determined from mathematical constraints, domain theory, and empirical observations.\vskip 0.01in}
\end{small}
\end{table}

To make entities explicit in the process modeling representation, a scientist specifies a set of generic entity types, each of which defines the properties of a class of objects.  Multiple instantiated versions of these generic entities can be included in a model.  For example, an ecologist could create a generic entity for birds, and then instantiate it as a sparrow or a hawk by changing the values of class properties.

\citet{todorovski05} report an initial formalism for specifying entities in process models that we build upon here in Table \ref{tab:ET-syn}.  The formal generic entity has a name and a set of properties.  Entities may contain both variables and parameters, where variables change in the context of the model and parameters do not.  Variables must have a name and a rule that determines how the effects of multiple processes are aggregated (\textit{e.g.}, summed, multiplied).  Parameters must have a name and an interval that delineates their possible values.  For both variables and parameters, there is an optional slot to provide a brief description.  In instantiated entities the variables are either associated with data or given initial values, the parameter values are assigned real values, and a field following the name indicates the parent generic entity.

To use an entity in a model, we need a notation that allows access to its fields.  We have chosen a dot notation that concatenates the variable or parameter name with the entity name. For example, to refer to the  chlorophyll \textit{a} concentration ($chla$) of an instantiated phytoplankton entity ($Phaeocystis$), we write $chla.Phaeocystis$.

\subsection{Processes}
Processes are the physical, chemical, or biological actions that drive change in dynamic models.  For example, growth is a biological process that occurs in many ecological models, whereas oxidation--reduction, photolysis, and sorption are examples of chemical and physical processes from biogeochemistry.  The process modeling framework employs two forms of processes: generic and instantiated.  Table \ref{tab:GP-syn} shows the syntax for generic processes, which define the basic properties of a class of processes.  Generic processes must include a name, a statement of which generic entities or entity types can be involved, and a set of equations.  The relates statement identifies unique entity roles in the process and the entity types that can fill those roles.  A generic process can also include a set of Boolean conditions that control whether the process is inactive.  For example, we could set the conditions so that the variable light must be positive for the photosyntheses process to occur ($light.environment > 0$).  Such statements turn processes on and off, making the model structurally dynamic. The set of generic processes are collected with the generic entities into one library.
\begin{table}[t]
\caption{The syntax for defining a generic process.} \label{tab:GP-syn}
\begin{small}
\tableline{\obeylines\leftskip 2pt\parskip 0 pt %
generic process \emph{name} \vskip 0.06in
~~relates $role_1$\{entity type, entity type\}$^a$ \vskip 0.03in
~~~~~~~~~~~$role_2$\{entity type, entity type\} \vskip 0.03in
~~parameters $<$set of process specific parameters$>$
~~equations $<$set of differential and algebraic equations$>^b$
~~conditions $<$set of boolean conditions$>^c$
} \tableline
{\scriptsize$^a$Each role may be played by multiple objects instantiated from different generic entities.\vskip 0.01in
$^b$Equations describe the effects of the process on the entities.\vskip 0.01in
$^c$Conditions control when the process is active.\vskip 0.01in}
\end{small}
\vskip 0.15in
\end{table}

The second form of process is the instantiated version of a generic process, which is bound to specific entities and has specific values for parameters.  Each instantiated process has a specific name and instantiated entities fill the roles in the equations.  The process naming scheme first identifies the process and then the entities the process affects.  For example, we might name an exponential growth process that operates on a phytoplankton entity `exp\_growth\_phyto'.

Table \ref{tab:GP-ex} illustrates the difference between instantiated and generic processes.  Both representations refer to an exponential death process.  The generic process specifies two roles that entities play in this process. The first role, played by the organism that dies, can be filled by either the phytoplankton or zooplankton entity types.  The equations indicate that these types must contain a variable called $conc$ and a parameter named $dr$. The second role can only be filled by the detritus entity type, which must also have a $conc$ variable.  In the instantiated process, the name of the instantiated entity replaces the role name in the equations.  Thus, the first role was filled by a phytoplankton entity named $Phaeocystis$, whereas the second role was filled by the detritus entity named $Detritus$.
\begin{table}[t]
\caption{An example of a generic process for exponential death and an instantiated version of the process.}
\label{tab:GP-ex}

\begin{small}
\tableline{\obeylines\leftskip 2pt\parskip 0 pt %
\textbf{Generic Process} \vskip 0.03in
generic process death\_exponential \vskip 0.01in
~~relates $S1$\{phytoplankton, zooplankton\}, $S2$\{detritus\} \vskip 0.01in
~~equations d[$conc.S1$,t,1] = -1 * $dr.S1$ * $conc.S1$ \vskip 0.01in
~~~~~~~~~~~~~~~d[$conc.S2$,t,1] = $dr.S1$ * $conc.S1$\vskip 0.06in
\textbf{Instantiated Process} \vskip 0.03in
process death\_exponential\_Phaeo \vskip 0.01in
~~equations d[$conc.Phaeocystis$,t,1] = -1 * $dr.Phaeocystis$ * $conc.Phaeocystis$ \vskip 0.01in
~~~~~~~~~~~~~~~d[$conc.Detritus$,t,1] = $dr.Phaeocystis$ * $conc.Phaeocystis$ \vskip 0.01in
} \tableline
\end{small}
\end{table}

A key distinction of this representation is that elements of the model equations that pertain to a process are grouped together.  This organization communicates how a process is formulated, what entities drive it, and how it affects particular entities.  This process centered association also facilitates the assembly of the system of equations for a model from these fragments.
\subsection{Modeling Procedure}\label{sec:modproc}
The modeling procedure described here resembles the traditional recipe found in many texts on quantitative modeling \citep[e.g.,][]{jorgensen01,grant97,gold77}, but it bears two distinctive features.  First, the procedure is organized as a search through a space of candidate models for those that explain a set of empirical observations.  Second, the method assembles candidate models from libraries of generic entities and processes.  To implement this approach, we must define the search space, describe a search mechanism, and specify criteria for selecting among alternative models.  In this paper we assume that the structural search is exhaustive, the parameter search is accomplished using existing algorithms such as gradient descent, and that the selection criteria include quantitative measures of goodness-of-fit (\emph{e.g.}, sum of squared errors, $r^2$) and the realism of the model structure.  These assumptions allow us to focus our discussion on the search space.

As described in the introduction, the search space is determined by alternative model structures and sets of parameters.  While this space can be quite large, our method of representing and developing models constrains it in two ways.  First, a model's structure comprises entities linked by processes.  The space of possible model structures is circumscribed by a domain-specific library of generic entities and processes thought to occur in the system, and specific entities to be related.  The generic types dictate how instantiated processes and entities can be combined.  Second, the entity types specify the possible ranges for each parameter, limiting the parameter space.  The range defined in the generic entity may be wide if we have little knowledge about the appropriate values, but additional knowledge, such as empirical estimates, can support a narrower range.

In the ideal case, one or more libraries of generic entities and processes will already exist for the domain being investigated.    In that case, the challenge is to select the most relevant library and adapt it as needed.  For example, it will be necessary to adapt a library when observations of the physical system suggest alternative entity definitions, or when the modeling objective is to determine the significance of a conjectured new process.

If no appropriate library exists, then an investigator must construct one to support the modeling task.  Although this requirement introduces an additional step to the modeling task, we believe that the advantages of using an explicit library of entities and processes outweigh the cost of its construction.  Specifically, the generic entities and processes define the search space.  From the generic definitions, the library determines the possible entities and processes the model can include, their possible structural relationships, alternative mathematical formulations, and the parameter ranges.  Further, to the degree that the library reflects existing domain knowledge, it also determines the space of theoretically plausible or realistic models.  In addition, a key idea in this approach is that these libraries form a repository of domain knowledge or theory that can be reused for different modeling projects.  We expect this to both expedite the modeling process and constrain model structures to reflect existing domain knowledge.  At the very least, we expect it will ease review of the domain theory and modeling assumptions used.

Given the space of plausible models, the investigator must search through it for models that explain the empirical observations.  This search has three steps:
\begin{enumerate}
\item bind entities to selected generic processes and combine these components into a model structure;
\item use estimation routine to identify parameters that generate predictions that best fit the observed variables; and
\item evaluate the resultant model's performance using a set of model selection criteria.
\end{enumerate}
In this way, multiple alternative models are constructed and evaluated.  This model search can be implemented by hand, but the formalism also allows both the structure and parameter search to be automated.  Automatic model induction is beyond the scope of this paper, but see \citet{langley02, langley00, langley03, george03, langleyAIM} for details on how this can be accomplished.

\subsection{Prometheus}\label{sec:prometheus}
To encourage the process modeling approach, we are developing a software environment called {\sc Prometheus} that is designed to support model building from conceptual development to evaluation, use, and publication \citep{sanchez03}.  This program has six principal features.  First, it provides an interactive way for a user to add and edit generic processes to a domain library.  Second, it allows manual construction of instantiated models from a domain library in an interactive and graphic environment.  For example, to add a process the user selects a generic process from the library, which opens a dialogue box that guides the user through binding the entity roles and parameters in the generic process to specific entities in the model.  The user can also specify the values of process level parameters in this dialogue.  Given an instantiated model, a third feature of the program is to assemble and simulate the model equations.  The user can then view the equations and graph the simulated trajectories.  The fourth feature is {\sc Prometheus}' ability to automatically search through candidate models derived from a library.  The program returns a reduced set of the models with trajectories that best match the observed data based on sum of squared errors.  A fifth feature is the ability to automatically revise an existing model.  Through an interactive dialogue, the user can ask {\sc Prometheus} to add or delete whole processes, and to revise parameters within processes.  \citet{asgharbeygi06} discuss this ability in more detail.  The final feature is a graphical representation of the model structure in which entity variables are linked through processes.  This representation of instantiated models provides another way for users to view and analyze their models (see Section \ref{sec:inst} for more details on the model diagrams).  Collectively, these features provide a comprehensive toolbox for process modeling.  

\section{Ross Sea Ecosystem Model}\label{sec:RossSeaModel}
In this section, we illustrate the process modeling approach with a version of the Ross Sea ecosystem embedded in the Ross Sea Coupled Ice And Ocean (CIAO) model \citep{arrigo03,worthen04}.  The original ecosystem model has six state variables, including \emph{P. antarctica} ($P_1$), diatoms ($P_2$), detritus ($D$), zooplankton ($Z$), nitrate ($NO_3$) and iron ($Fe$).  In CIAO, this ecosystem model is coupled to a modified version of the Princeton Oceanographic Model, which is a physical ocean model that predicts the three dimensional structure of temperature, salinity, and velocity in response to both surface and lateral boundary conditions of heat, salt, and momentum.  \citet{arrigo03} used CIAO to investigate factors controlling phytoplankton production and taxonomic composition in the Ross Sea, and \citet{worthen04} used it to explore the interannual variability of primary production. More recently, \citet{tagliabue05} added more detailed mechanisms for iron cycling, the carbonate system, and the air-sea CO$_2$ exchange.  They used this enhanced model to investigate the sensitivity of the ecosystem dynamics to changes in taxon-specific nutrient utilization parameters.  They also explored the impact of differing iron fertilization regimes on the ability of the system to sequester carbon \citep{arrigo05}.  In addition, \citet{asgharbeygi06} employed a modified version of the ecosystem model to demonstrate the usefulness of new inductive tools for model revision.

To more clearly demonstrate the process modeling approach, we simplified the Ross Sea ecosystem model in two ways.  First, we extracted the ecosystem model from CIAO, and focused on activity near the ocean surface.  In this simplified model, there are three driving variables:  daily mean photosynthetically usable radiation at the ocean surface ($PUR_e$) and center ($PUR_c$) of the modeled volume and water temperature ($T_{H_2O}$).  Second, we represent phytoplankton with one state-variable ($P$) rather than two.  The variables $C$, $Z$, and $D$ represent the concentration of carbon in the three entities, and $NO_3$ and $Fe$ represent the concentration of nitrate and iron, respectively.
\subsection{Differential and Algebraic Equation Representation}
As is common in ecological modeling, the original Ross Sea ecosystem model was originally formulated as a set of ordinary differential and algebraic equations, which were implemented as a Fortran program.  The modified differential equations for the five state-variables are
\begin{eqnarray}
\frac{dP}{dt}&=&((1-e_P)\mu-x_P-\omega_P)P-gZ  \label{eq:P},\\
\frac{dZ}{dt}&=&(\gamma_Z g-x_Z-r_Z)Z  \label{eq:Z},\\
\frac{dD}{dt}&=&(1-\beta)[(1-\gamma_Z)gZ+x_PP+x_ZZ] -(r_D+\omega_D)D\label{eq:D},\\
\frac{dNO_3}{dt}&=&-\phi_{(P,N/C)}\mu P \label{eq:NO3}, \text{ and}\\
\frac{dFe}{dt}&=& \phi_{(D,Fe/C)}r_DD-\phi_{(P,Fe/N)}\mu P.  \label{eq:Fe}
\end{eqnarray}

In equation \ref{eq:P}, the balance between the rates of growth ($\mu$), death ($x_P$), exudation ($e_P$), grazing ($g$), and sinking ($\omega_P$) determine the change in phytoplankton concentration ($P$).  The product of the grazing rate ($g$), the zooplankton assimilation efficiency ($\gamma_Z$), and the phytoplankton being grazed, as well as the zooplankton death ($x_Z$) and respiration ($r_Z$) rates determine the changes in zooplankton ($Z$) abundance as shown in equation \ref{eq:Z}.  Additions to the detrital ($D$) pool, modeled in equation \ref{eq:D}, include the particulate fractions ($1-\beta$) of unassimilated grazing (\emph{i.e.}, fecal pellets) and dead phytoplankton and zooplankton.  Losses are from remineralization ($r_D$) and sinking ($\omega_D$).  Phytoplankton nitrate uptake is assumed to be proportional to the magnitude of phytoplankton growth ($\mu P$), and determines the Nitrate ($NO_3$) concentration as equation \ref{eq:NO3} illustrates. We then multiply this quantity by the phytoplankton nitrogen to carbon ratio ($\phi_{(P,N/C)}$).  There are no significant nitrate inputs during the growing season.  In equation \ref{eq:Fe}, iron concentration ($Fe$) is a function of detrital remineralization (adjusted by the Fe/C ratio, $\phi_{(D,Fe/C)}$) and phytoplankton production multiplied by the phytoplankton Fe/C ratio $(\phi_{(P,Fe/C)}$).

Phytoplankton growth rate ($\mu$) and zooplankton grazing rate ($g$) are variables in this model.  We formulated the phytoplankton growth rate as
\begin{equation}\label{eq:growth}
\mu=\mu0_{max}\cdot e^{(0.06933 \cdot T_{H_2O})}\cdot \min(ik\_lim,n\_lim,fe\_lim),
\end{equation}
where $\mu0_{max}$ is the maximum unlimited growth rate at 0$^\circ$C and $T_{H_2O}$ is the water temperature.  Light and nutrients limit growth as if they were substitutable resources.  That is, only the minimum value of these resources retards the maximum growth rate.  We calculated light limitation ($ik\_lim$) as
\begin{subequations} \begin{align}\label{eq:iklim}
ik\_lim&=a-e^{(-PUR_c/ik)}\cdot e^{(-PUR_c/photoinhib)}\\
ik&=ik_{max} / \big( a+2.50 \cdot e^{\big(PUR_e\cdot e^{(1.089-1.12\cdot \log_{10}(ik_{max}))}\big)}\big)
\end{align} \end{subequations}
where $a$ is a fitting parameter, $PUR_c$ and $PUR_e$ are the mean daily photosythetically usable radiation ($\mu$mol photons m$^{-2}$ s$^{-1}$) in the center of the mixed ocean layer and at the mixed layer edges, respectively, $photoinhib$ represents the negative impact of high light on phytoplankton growth, and $ik_{max}$ is the upper limit of $ik$.  Nitrate ($n\_lim$) and iron ($fe\_lim$) limitation are modeled with Monod functions, giving
\begin{subequations}\label{eq:nut-lim}
\begin{align}
n\_lim&=\frac{NO_3}{NO_3+k_{NO_3}}\label{eq:nlim} \textrm{ and}\\
fe\_lim&=\frac{Fe}{Fe+k_{Fe}}\label{eq:felim},
\end{align} \end{subequations}
where $k_{NO_3}$ and $k_{Fe}$ are the half-saturation constants for nitrate and iron, respectively.

Grazing assumes a Holling Type II relationship,
\begin{subequations}\label{eq:graze} \begin{align}
g&=\frac{g_{max}\cdot\eta}{\eta + g_{cap}} \label{eq:g}\\
\eta&=P-(biomin + g_{lim}), \label{eq:eta}
\end{align} \end{subequations}
where $g_{max}$ is the maximum grazing rate of $Z$ on $P$, $g_{cap}$ is the capture rate, and $\eta$ represents the phytoplankton concentration susceptible to grazing.  Here, $biomin$ is a modeling measure to ensure that phytoplankton concentrations remain positive, and $g_{lim}$ represents the biological limitation of zooplankton grazing on phytoplankton.

Representing the model as a system of differential and algebraic equations has advantages, in that it clearly identifies the model variables and the full equations such that translating them into computer code for simulation would be relatively trivial.  The differential equations imply the entities Phytoplankton, Zooplankton, Detritus, Nitrogen, and Iron.  Parameters associated with an entity have its state-variable as a subscript.  Although the entities are implicit, the identities of the causal processes are obscured.  Close inspection of the equations reveals that certain elements correspond to specific physical, chemical, and biological processes.  Unfortunately, no specific markers for the processes exist, and in some cases the processes include elements from multiple equations.
\subsection{Process Modeling Representation}
Here, we illustrate the process modeling approach and provide a corresponding representation of the Ross Sea ecosystem.  We designed the modeling procedure to facilitate the construction of new models, but the task here differs somewhat in that we are translating an existing model.  This presents additional challenges that highlight important features of the approach.

As discussed in Section \ref{sec:modproc}, the first step of the process modeling procedure is to create or to select a library of entity types and generic processes. In this case, we need to generate a new library; therefore, we must identify the entities, entity types, processes and generic processes we expect to be significant in the model.  Here, we infer these items from the existing equations.  We conclude this section by demonstrating how to build the instantiated model of the Ross Sea ecosystem using the library.
\subsubsection{Entities}
As already stated, entities are implicit in the original Ross Sea model; We can infer them from the equations by examining the variables and parameters.  Each of the variables in the left hand side of the differential equations (\ref{eq:P}--\ref{eq:Fe}) describe separate entities that are associated with parameters through subscripts.  In addition, we associate the driving variables ($PUR_e$, $PUR_c$ and $T_{H_2O}$) with a separate entity we label \textit{Environment}.  Further, we assign the variables characterizing phytoplankton growth rate ($\mu$) and light and nutrient growth rate limitations to the phytoplankton entity.

The grazing rate ($g$) presents complications.  It forms from the \textit{interaction} of two entities, and it does not necessarily belong to either one.  We might associate it with a grazing process, but this is problematic.  We show later that we need this variable to operate in multiple processes, but in our framework variables and parameters associated with a process cannot be accessed by another process.  We could create a new entity to house this variable, but for now we associate it with \textit{Environment}.

The next task is to determine if these entities are unique types, or if one or more might be derived from the same general class.  To make this decision, we examined the parameters associated with each entity along with the roles the entities play in the model. Since nitrate and iron are both nutrients and operate similarly in the model, we decided to create generic entity called \textit{Nutrient}.  The remaining entities have unique types.

Tables \ref{tab:entity-1} and \ref{tab:entity-2} define the entity types (\textit{Phytoplankton}, \textit{Zooplankton}, \textit{Detritus}, \textit{Nutrient}, and \textit{Environment}) using the process modeling formalism.  Presently, \textit{Phytoplankton} and \textit{Environment} are the only entity types with more than one variable, which reflects the central roles they play in the Ross Sea model.  In addition to the driving variables and the grazing rate, we associated the parameter $\beta$ with \textit{Environment}.  This parameter determines the soluble and insoluble fractions of organic matter, and supplants modeling the detailed processes of adsorption, flocking, and dissolution.  While we designed these five entity types to support the Ross Sea model, they could easily be modified to support other aquatic ecosystem models.
\begin{table}[t]
\caption{Entity types for the Ross Sea process model.} \label{tab:entity-1}
\begin{small}
\tableline{\obeylines\leftskip 2pt\parskip 0 pt %
entity \textbf{Phytoplankton} \vskip 0.01in %
~~description This entity is a composite of the phytoplankton species. \vskip 0.01in %
~~variables \vskip 0.01in %
~~~~~~$conc$ \{sum\}, carbon concentration of phytoplankton  \vskip 0.01in%
~~~~~~$\mu$ \{sum\}, realized growth rate \vskip 0.01in%
~~~~~~$ik\_lim$ \{sum\}, light limitation of maximum growth rate \vskip 0.01in%
~~~~~~$n\_lim$ \{sum\}, nitrogen limitation of maximum growth rate \vskip 0.01in%
~~~~~~$fe\_lim$ \{sum\}, iron limitation of maximum growth rate \vskip 0.01in%
~~parameters \vskip 0.01in %
~~~~~~$\mu0_{max}$ [0,1], maximum unlimited phytoplankton growth rate at 0$^\circ$C \vskip 0.01in%
~~~~~~$e$ [0,1], exudation rate \vskip 0.01in%
~~~~~~$x$ [0,1], death rate \vskip 0.01in%
~~~~~~$ik_{max}$ [0,$\infty$], upper limit of $ik$ \vskip 0.01in%
~~~~~~$\omega$ [0,1], sinking rate  \vskip 0.01in%
~~~~~~$biomin$ [0,$\infty$], minimum carbon concentration \vskip 0.01in
~~~~~~$bioinhib$ [0,$\infty$], photoinhibition of phytoplankton growth \vskip 0.01in
~~~~~~$\phi_{(N/C)}$ [0,1], ratio of nitrogen to carbon in phytoplankton \vskip 0.01in
~~~~~~$\phi_{(Fe/C)}$ [0,1], ratio of iron to carbon in phytoplankton
\vskip 0.2in

entity \textbf{Zooplankton} \vskip 0.01in
~~description This entity is a composite of the zooplankton species \vskip 0.01in
~~variables \vskip 0.01in
~~~~~~$conc$ \{sum\}, carbon concentration of zooplankton \vskip 0.01in
~~parameters \vskip 0.01in
~~~~~~$\gamma$ [0,1], assimilation efficiency \vskip 0.01in%
~~~~~~$x$ [0,1], death rate \vskip 0.01in%
~~~~~~$r$ [0,1], respiration rate \vskip 0.01in%
~~~~~~$\omega$  [0,1], sinking rate  \vskip 0.01in%
~~~~~~$g_{max}$ [0,1], maximum grazing rate    \vskip 0.01in
~~~~~~$g_{lim}$ [0,$\infty$], grazing limitation    \vskip 0.01in
~~~~~~$g_{cap}$ [0,$\infty$], grazing capacity  \vskip 0.01in
} \tableline
\end{small}
\vskip -0.15in
\end{table}

\begin{table}[t]
\caption{Entity types for the Ross Sea process model (continued).} \label{tab:entity-2}
\begin{small}
\tableline{\obeylines\leftskip 2pt\parskip 0 pt %
entity \textbf{Detritus} \vskip 0.01 in %
~~description suspended particulate organic matter \vskip 0.01in %
~~variables \vskip 0.01in %
~~~~~~$conc$ \{sum\}, carbon concentration of detritus \vskip 0.01in%
~~parameters \vskip 0.01in %
~~~~~~$r$ [0,1], remineralization rate of detritus \vskip 0.01in%
~~~~~~$\omega$ [0,1], sinking rate  \vskip 0.01in%
~~~~~~$\phi_{(N/C)}$ [0,1], ratio of nitrogen to carbon in detritus 
\vskip 0.15in

entity \textbf{Nutrient} \vskip 0.01 in %
~~description dissolved nutrients in the water column %
\vskip 0.01in %
~~variables \vskip 0.01in %
~~~~~~$conc$ \{sum\}, dissolved nitrate concentration
\vskip 0.15in

entity \textbf{Environment} \vskip 0.01in %
~~description the system environment
~~variables \vskip 0.01in %
~~~~~~$T_{H_2O}$ \{sum\}, sea water temperature \vskip 0.01in %
~~~~~~$PUR_c$ \{sum\}, average daily light intensity at the center of the modeled volume \vskip 0.01in %
~~~~~~$PUR_e$ \{sum\}, average daily light intensity at the top edge of the modeled volume \vskip 0.01in %
~~~~~~$g$ \{sum\}, rate of zooplankton grazing on phytoplankton \vskip 0.01in
~~parameters \vskip 0.01in %
~~~~~~$\beta$ [0,1], proportion of matter that becomes soluble \vskip 0.01in}
\tableline
\end{small}
\vskip -0.15in
\end{table}

\subsubsection{Processes}
To generate a generic process library we must know what physical, chemical, and biological processes might be relevant in the target model.  Again, we analyzed equations 1--9 to identify the processes required in our library.  Our knowledge of the system, aquatic ecosystems in general, and ecological theory guided this analysis and the construction of the library.  We present the resulting generic process library in Tables \ref{tab:rsGPL-1} and \ref{tab:rsGPL-2}.
\begin{table}[t]
\caption{Generic process library used to construct a revised Ross Sea ecosystem model.} \label{tab:rsGPL-1}
\begin{small}
\tableline{\obeylines\leftskip 2pt\parskip 0 pt %
library RossSea \vskip 0.06in
generic process growth\_exp \vskip 0.01in
~~relates $P\{Phytoplankton\}, NO_3\{Nutrient\}, Fe\{Nutrient\}$ \vskip 0.01in  
~~equations $d[conc.P,t,1]= (1-e.P) * \mu.P * conc.P$ \vskip 0.01in
\hskip 0.79in $d[conc.NO_3,t,1]= -1 * \phi_{(N/C)}.P * \mu.P * conc.P$ \vskip 0.01in
\hskip 0.79in $d[conc.Fe,t,1] = -1 * \phi_{(Fe/C)}.P * \mu.P * conc.P$ \vskip 0.08in

generic process death\_exp \vskip 0.01in
~~relates $S\{Phytoplankton, Zooplankton\}, D\{Detritus\}, E\{Environment\}$ \vskip 0.01in
~~equations $d[conc.S,t,1]= -1 * x.S * conc.S$ \vskip 0.01in
\hskip 0.79in $d[conc.D,t,1]= (1-\beta.E) x.S * conc.S$ \vskip 0.08in

generic process grazing \vskip 0.01in
~~relates $P\{Phytoplankton\}, Z\{Zooplankton\}, D\{Detritus\}, E\{Environment\}$ \vskip 0.01in
~~equations $d[conc.P,t,1]= -1 * g.E * conc.Z$ \vskip 0.01in
\hskip 0.79in $d[Z,t,1]= \gamma.Z * g.E * conc.Z$ \vskip 0.01in
\hskip 0.79in $d[D,t,1]= (1-\beta.E) * (1-\gamma.Z) * g.E * conc.Z$ \vskip 0.08in
generic process respiration  \vskip 0.01in
~~relates $Z\{Zooplankton\}$ \vskip 0.01in
~~equations $d[conc.Z,t,1]= -1 * r.Z * conc.Z$ \skip 0.08in

generic process remineralization\vskip 0.01in
~~relates $D\{Detritus\}, N\{Nutirent\}$\vskip 0.01in
~~equations $d[conc.D,t,1] = -1 * r.D * conc.D$  \vskip 0.01in
\hskip 0.79in $d[conc.N,t,1] = \phi_{(N/C)}.D * r.D * conc.D$ \vskip 0.01in

generic process sinking\vskip 0.01in
~~relates $V\{Phytoplankton, Detritus\}$ \vskip 0.01in
~~equations $d[conc.V,t,1] = -1 * \omega.V * conc.V$  \vskip 0.08in

generic process growth\_rate \vskip 0.01in
~~relates $E\{Environment\}, P\{Phytoplankton\}$   \vskip 0.01in
~~equations $\mu.P = (\mu0_{max}.P * exp(0.0633*T_{H_20}.E))$ * \vskip 0.01in
\hskip 0.8in $min(ik\_lim.P, n\_lim.P, fe\_lim.P)$
} \tableline
\end{small}
\vskip -0.15in
\end{table}
\begin{table}[t]
\caption{Generic process library used to construct a revised Ross Sea ecosystem model (Continued).} \label{tab:rsGPL-2}
\begin{small}
\tableline{\obeylines\leftskip 2pt\parskip 0 pt %
generic process growth\_limitation\_light\_w\_photoinhibition\vskip 0.01in
~~relates  $P\{Phytoplankton\}, E\{Environment\}$ \vskip 0.01in
~~parameters $a$ [0,2], fitting parameter \vskip 0.01in
~~equations $ik\_lim.P = a - exp(-PUR_c.E/$ \vskip 0.01in
\hskip 0.78in $(ik_{max}.P/(a + 2.50 * exp(PUR_e.E*exp(1.089-1.12*log_{10}(ik_{max}.P))))))*$ \vskip 0.01in
\hskip 0.78in $exp(-PUR_c.E / photoinhib.P)$ \vskip 0.08in

generic process monod\_limitation\vskip 0.01in
~~variables $V\{Nutrient\}$ \vskip 0.01in
~~parameters $k$, half-saturation constant
~~equations $m\_lim.V = con.V / (conc.V + k)$ \vskip 0.08in

generic process grazing\_rate\vskip 0.01in
~~variables $P\{Phytoplankton\},Z\{Zooplankton\}, E\{Environment\}$ \vskip 0.01in
~~equations $g.E = (g_{max}.Z * (conc.P - (biomin.P+ g_{lim}.Z))) / $\vskip 0.01in
\hskip 0.78in $(g_{cap}.Z + (conc.P - (biomin.P+ g_{lim}.Z)))$
} \tableline
\end{small}
\vskip -0.1in
\end{table}

In our analysis, we found that parameters in the original model guided the identification of processes because they were originally selected to represent particular aspects of the real system.  To use these tags, we fully expanded the differential equations and marked equation elements that seemed to correspond to processes.  For example, we rewrote and labeled equation \ref{eq:P} as
\begin{equation}
\frac{dP}{dt}= \underbrace{\mu P}_\textrm{Growth} -\underbrace{e\mu P}_\textrm{Exudation} - \underbrace{x_PP}_\textrm{Death} - \underbrace{\omega_PP}_\textrm{Sinking} - \underbrace{gZ}_\textrm{Grazing}.  \label{eq:P2}
\end{equation}

We initially identified five processes in this equation: growth, exudation, death, sinking, and grazing.  However, growth and exudation are not independent in the equation formulation because exudation is a constant proportion of the amount of new growth ($\mu P$).  When we applied the same procedure to the remaining differential equations we found that what we initially identified as a nutrient uptake process in equations (\ref{eq:NO3}) and (\ref{eq:Fe}) ($-\phi_{(P,C/N)}\mu P$ and $-\phi_{(P,C/Fe)}\mu P$, respectively) was also formulated as a direct proportion of the amount grown.  Therefore, we treated exudation and nutrient uptake as sub-processes of growth, and bundled them into one phytoplankton growth process. We made this decision to improve the understandability and causal flow of the process model.

Although we divided the differential equations into different processes, we generally mapped the algebraic equations (\ref{eq:growth}-\ref{eq:graze}) into separate processes (Table~ \ref{tab:rsGPL-2}).  While bundling the calculation of the grazing rate (\ref{eq:graze}) into the grazing process, and the equations for growth rate (\ref{eq:growth}) with its nutrient and light limitations (\ref{eq:iklim}, \ref{eq:nlim}, \ref{eq:felim}) is conceptually reasonable, we chose to disaggregate these processes so that we could track them individually.  In addition, segregating these limitation processes lets us easily construct alternative processes with a different formulation.  For example, we could construct an alternative light limitation process that does not include the photoinhibition term. In total, we encoded 11 generic processes that could operate in the Ross Sea ecosystem model.

The process modeling formalism builds a new layer of information into the model that reflects additional modeling decisions regarding entities and processes.  The growth process provides a good example.  As we described earlier, we chose to include exudation and nutrient uptake in the phytoplankton growth process because they were not modeled as independent processes.  However, this decision obscures the identity of these subprocesses.  If the modeling task required that we maintain their identities, we could define a new variable associated with the phytoplankton entity to represent the instantaneous amount of new phytoplankton growth ($new\_growth.P$).  With this new variable, we show how we could formulate separate processes for exudation and the nutrient uptake in Table \ref{tab:RossSea-alt}.  This technique of identifying \emph{hidden variables} is especially useful when multiple processes that might be considered sub-processes impact the product of a process, or when we want a series of sub-processes to be explicit.  These additional decisions make it a non-trivial task to translate from equations to entities and processes or to construct an initial library, while assembling the equations from entities and processes is simple.
\begin{table}[b]
\caption{Alternative processes to explicitly model growth, exudation, and nutrient uptake} \label{tab:RossSea-alt}
\begin{small}
\tableline{\obeylines\leftskip 2pt\parskip 0 pt %

process growth\_phyto \vskip 0.01in %
~~equations $d[conc.P,t,1] = \mu * conc.P $ \vskip 0.01in
\hskip 0.79in $new\_growth.P = \mu * conc.P$\vskip 0.08in

process exudation \vskip 0.01in %
~~equations $d[conc.P,t,1] = -1 * e.P * new\_growth.P$\vskip 0.08in

process growth\_nutrient\_uptake\_NO3 \vskip 0.01in %
~~equations $d[conc.NO_3,t,1] = -\phi_{(N/C)}.P * new\_growth.P$ \vskip 0.08in %

process growth\_nutrient\_uptake\_Fe \vskip 0.01in %
~~equations $d[conc.Fe,t,1] = -\phi_{(Fe/C)}.P * new\_growth.P$

\vskip 0.03in} \tableline
\end{small}
\end{table}

\subsubsection{Instantiated Model} \label{sec:inst}
Given a library of generic entities and processes, we instantiate candidate models using two steps.  In the first step, we specify the instantiated entities.  For the Ross Sea model, we instantiated one entity for Phytoplankton ($P$), Zooplankton ($Z$), Detritus ($D$), and Environment ($E$).  We also instantiated two Nutrient entities named $NO3$ and $Fe$.  In the second step, we bind specific instances of generic processes to members of the entity set.  Then, using the modeling procedure described in Section \ref{sec:modproc}, we can consider multiple models with different structures.  With the library in Tables \ref{tab:rsGPL-1} and \ref{tab:rsGPL-2}, one of the possible candidate models will have the same structure as the original Ross Sea model, while the remaining candidates will have a subset of the original model's structure.  We could expand the number of candidate models and increase the number of processes in the model by adding new generic processes or alternative formulations of existing ones to the library.

We present a portion of the the original Ross Sea ecosystem model specified with this formalism in Table \ref{tab:RS-entity}.  The instantiated model begins with the model name, after which come the instantiated entities and their types.\footnote{In this example we retain the parameter symbols, rather than providing parameter values}  The next line declares the observed variables (\textit{i.e.}, those for which data are available).  The set of instantiated processes follows.  The complete instantiated model (not shown) contains thirteen processes.  Although this view of the model is not as concise as the final set of equations, it conveys more information about the model structure.
\begin{table}[t]
\caption{A select portion of the instantiated Ross Sea Ecosystem model represented in the process modeling formalism.  Parameter values are not shown.}
\label{tab:RS-entity}
\begin{small}
\tableline{\obeylines\leftskip 2pt\parskip 0 pt %

model RossSea \vskip 0.06in %
entities $P\{Phytoplankton\}, Z\{Zooplankton\}, D\{Detritus\}, NO_3\{Nitrogen\},$ \vskip 0.01in
\hskip 0.5in $Fe\{Iron\}, E\{Environment\}$ \vskip 0.08in

observable $conc.P$, $conc.NO_3$\vskip 0.08in
exogenous $T_{H_2O}.E$, $PUR_c.E$, $PUR_e.E$\vskip 0.08in

process growth\_phyto \vskip 0.01in %
~~equations $d[conc.P,t,1] = (1-e.P) * \mu.P * conc.P$ \vskip 0.01in %
\hskip 0.79in $d[conc.NO_3,t,1] = -\phi_{N/C}.P * \mu.P * conc.P$ \vskip 0.01in %
\hskip 0.79in $d[conc.Fe,t,1] = -\phi_{Fe/C}.P * \mu.P * conc.P$ \vskip 0.08in

process death\_phyto \vskip 0.01in %
~~equations $d[conc.P,t,1] = -x.P * conc.P$ \vskip 0.01in %
\hskip 0.79in $d[conc.D,t,1] = (1-\beta.E) * x.P* conc.P$ \vskip 0.08in

process death\_zoo \vskip 0.01in %
~~equations $d[conc.Z,t,1] = -x.Z * conc.Z$ \vskip 0.01in %
\hskip 0.79in$d[conc.D,t,1] = (1-\beta.E) * x.Z * conc.Z$ \vskip 0.08in

process grazing \vskip 0.01in %
~~equations $d[conc.P,t,1] = -g.E * conc.zoo$ \vskip 0.01in %
\hskip 0.79in $d[conc.Z,t,1] = \gamma.Z * g.E * conc.Z$ \vskip 0.01in %
\hskip 0.79in $d[conc.D,t,1] = (1-\beta.E) * (1-\gamma.Z) * g.E * conc.Z$ \vskip 0.08in

process remineralization\_iron \vskip 0.01in %
~~equations $d[conc.D,t,1] = -r.D * conc.D$ \vskip 0.01in %
\hskip 0.79in $d[conc.Fe,t,1] = \phi_{Fe/C}.D * r.D * conc.D$ \vskip 0.01in
} \tableline
\end{small}

\vskip -0.15in
\end{table}

\begin{figure}[t]
\includegraphics[scale=.55]{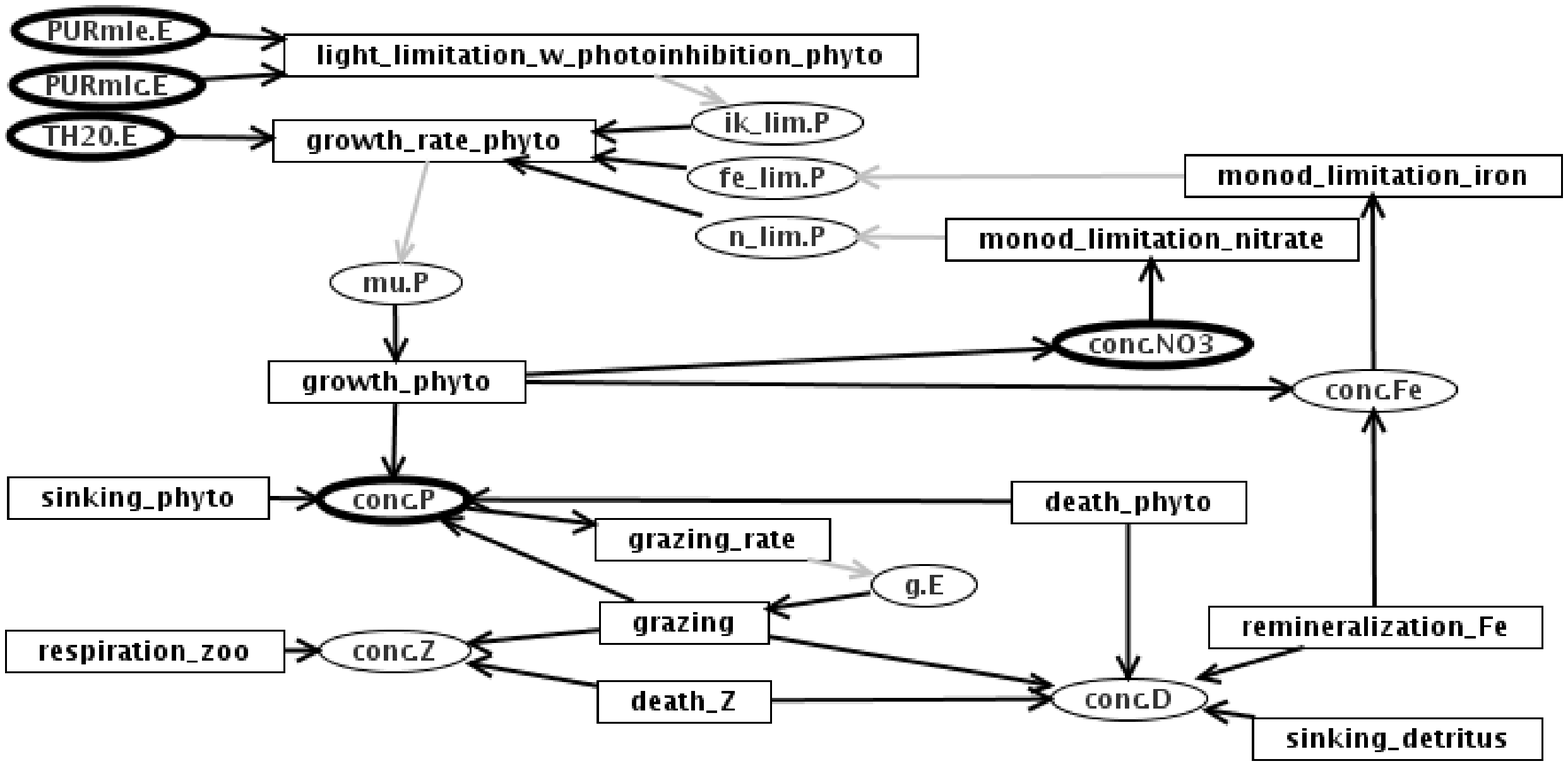}
\caption{Graph of the process representation of the Ross Sea ecosystem model.  Variables are represented as ovals and processes as rectangles.  Arrows indicate direct causal influence between processes and entity variables.  Light arrows indicate an algebraic relations, while dark arrows signify differential equation-based relationship.  Observed and driving variables are highlighted with a bold outline.}
\label{fig:RossSea-pfm}
\end{figure}

Figure \ref{fig:RossSea-pfm} shows the causal graph of the full Ross Sea process model drawn by {\sc Prometheus}.  Model variables are represented as ovals, processes are rectangles, and influence is shown by arrows.  This diagram shows explicitly which processes and variables are included in the model, how they are directly linked, and where indirect connections occur.  The ontological commitments forbid the direct connection of two variables.  Instead, the influence must be mapped through specific processes.  This lets us explicitly map two or more minimum causal paths ($variable_1 \rightarrow$ process $\rightarrow variable_2$) between two variables.  Inspection of the diagram reveals three feedback loops.  The first, $conc.P \rightarrow$ grazing\_rate $\rightarrow g.E \rightarrow$ grazing $\rightarrow conc.P$, shows the density dependence of grazing on the phytoplankton.  Two additional loops show the feedback of phytoplankton growth into itself through the two nutrient entities (\textit{e.g.}, growth\_Phyto $\rightarrow conc.NO3 \rightarrow$ monod\_limitation\_nitrate $\rightarrow n\_lim.P \rightarrow$ growth\_rate\_phyto $\rightarrow \mu.P \rightarrow$ growth\_Phyto).  Feedback loops are difficult to see in either the equation or process model representations alone, but they stand out in the network representation.

Along with recovering the original model, the process modeling approach lets users explore alternative model structures with relative ease.  Exhaustively searching the library, we would also find many subsets of the original model that may fit the observed data.  Consider the alternative structure suggested by \citeauthor{tagliabue03}'s (2003) finding that zooplankton may not be a significant factor in the plankton dynamics of the Ross Sea.  To explore this possibility, we could instantiate a model that excludes the zooplankton entity and its related processes.  Within our framework, we would then compare the performance of this model to the original one to help determine which is most appropriate.  If we wanted to expand the model so that it differentiates between \textit{P. antarctica} and the diatom phytoplankton, we could simply instantiate the phytoplankton entity type twice (\textit{e.g.}, replace $P\{Phytoplankton\}$ in the entity declaration line with $phaeo\{Phytoplankton\}$ and $diatoms\{Phytoplankton\}$) and then instantiate the processes associated with phytoplankton a second time, assigning one set to each phytoplankton group.  If the scientists do not believe the same processes impact the two groups, they can search through alternative process instantiations.  Again, we can add new generic processes or alternative formulations of generic processes to the library and implement them with little difficulty.  The ease with which users can develop and evaluate alternative model structures is one of the primary advantages of the process model formulation.
\section{Related Research}
The process modeling framework builds upon and relates to ideas already present in the literature.  Here, we characterize the work's relevance and novelty.

Modeling the processes that generate system dynamics is not a new idea in ecology.  In their textbook, \citet{jorgensen01} make a strong commitment to modeling specific processes.  In Chapter 3, the authors describe several mathematical formulations of physical, chemical, and biological processes that commonly occur in ecological models.  \citet{gurney98} also discuss the importance of processes in modeling ecological dynamics, stating ``formulation of a dynamic model always starts by identifying the fundamental processes in the system under investigation.''  However, both texts switch to sets of equations after processes are considered during model formulation.  An additional emphasis on processes in ecological modeling appears in the growing use of  process-based models \citep[e.g.,][]{landsberg97, simioni00, mcmurtrie85, brugnach05, melillo93, reynolds99, galbraith80}. With an emphasis on processes, these models resemble those built with the approach described in this paper.  \citet{brugnach05} shows clearly that process-based models are conceptualized as a network of data flowing between processes that operate on the data.  However, model actors or entities are not clearly delineated in this approach.

In an effort to create a general theory of modeling and simulation, \citet{zeigler74} presents a framework that is very similar to the representation we describe.   He describes models using three elements: components, descriptive variables, and component interactions.  Components are the parts of the system from which the model is constructed.  Variables (and parameters) characterize the properties of the individual components, and collectively portray the system's behavior.  Component interactions are the relationships linking components.  From this description, it is seems that entities in our framework captures both Zeigler's components and descriptive variables.  Our processes are a subset of Zeigler's component interactions.  The difference between our representation and Zeigler's approach stems from his desire to construct a general theory of modeling.  In contrast, we focus on constructing continuous-time, simulation models of mechanisms.

Our conceptual framework is perhaps most similar to one described by \citet{machamer00} in their discussion of mechanisms in biology.  They claim that mechanisms are composed of entities and activities.  As in our framework, entities are the actors in a system, both creating and being affected by activities, which produce change in the system.  Their activities seem equivalent to our processes, although the authors discuss and apply the framework to scientific reasoning in neurobiology and molecular biology.  They do not show how it relates to the mathematical models that arise in ecology, and they do not develop their ideas into a clear working methodology.

As in the process modeling approach, some ecological research generates and tests alternative possibilities.  At the broadest level, if we consider each candidate model structure as a hypothesis of the system's structure and function or an alternative explanation for observed behavior, then searching the space of hypotheses is an extreme form of evaluating multiple alternative hypotheses \citep[e.g.,][]{loehle87b, carpenter98, hilborn97}.  More closely related is the work of \citet{jost01}, who manually fit models with alternative formulations of the predation process to several predator--prey data sets.  Their objective was to determine whether predation was best explained by predator density-dependent or -independent processes.  Although they compared alternative processes, they view their modeling approach as a statistical fitting and selection of nonlinear models to population data, a methodology gaining prominence in population ecology \citep[e.g.,][]{jost00,jost01,hilborn97,carpenter94}.  However, their approach lacks an explicit notion of modeling as search.

The most closely related research is the approach to model revision introduced by \citet{reynolds99}.  They describe a multi-criteria assessment technique for evaluating process-based models that is used in an iterative modeling cycle to guide revision of both model structure and parameter selection.  Like the modeling approach discussed in this paper, the goal is to discover better model formulations.  Two important differences stem from the fact that their approach does not provide a way to specify domain theory \emph{a priori}.  First, their procedure only allows revision of an existing model; I t cannot guide search for the initial model.  Second, the methodology detects and generally locates model deficiencies, but it cannot suggest specific model revisions to make.  The scientist must manually revise the model and then reapply the multi-criteria assessment.

\section{Directions for Future Work}
As the Ross Sea model shows, the approach described in this paper can be an effective tool for constructing, organizing, and communicating complicated models.  However, there remain several ways in which the approach can be advanced.  In particular, we see three ways of extending our approach to improve its utility for modeling environmental phenomena.

The first extension involves implementing a hierarchical organization of both entities and processes.   Part of the complexity of environmental systems is that their components operate and interact with each other at multiple scales of space and time.  Hierarchical entities and processes are one way to capture this complexity that matches how biologist and ecologists tend to think about them \citep{ahl96, allen82, oneill86}. For example, organisms are classified using a taxonomic hierarchy.  The White-throated Sparrow (\emph{Zonotrichia albicollis}) is in the family Emberizidae, which is in the class Aves of the Animal kingdom \citep{NGS99}.  Members of each taxonomic level share certain properties that are used to classify them.  We may also use hierarchical entities to encode spatial phenomena.  We envision using a set of entities to represent different physical locations, where each ``spatial entity'' contains relevant entities and processes for that space.  For example, in a lake ecosystem model we might represent the areas above and below the thermocline as two distinct entities.  Finally, hierarchical processes would let us explicitly represent exudation and nutrient uptake in the Ross Sea ecosystem model as subprocesses of growth. We have already begun work on developing ways to represent and use such hierarchical processes \citep{todorovski05}.

The second extension addresses the criteria for selecting models.  Information about ecological systems is typically heterogeneous; ecologists generally know more about some parts of the system then others.  For example, they may have continuous-time observations for some system variables but only know the appropriate ranges or general trajectory shapes for others.  Scientists often evaluate the plausibility of simulation models using all of these criteria \citep[e.g.,][]{arrigo03}.   At present, {\sc Prometheus} only provides information about standard goodness-of-fit measures for the variables with continuous-time data.  Although we may use this additional information to evaluate the models manually, we may also be able to encode the information and use it to guide automated search and selection of appropriate models.

Finally, this modeling methodology creates new possibilities for analyzing system behavior.  Sensitivity analysis is commonly used in ecological modeling to determine the effect of varying one or more parameters on the model behavior.  However, we expect that performing a process-level sensitivity analysis will lead to increased understanding of system dynamics because it will reveal which processes are primarily responsible for dynamics at a given time.  This information will be useful for guiding additional experimental work, as well as assisting environmental assessment and management.  Using the more common process-based ecological modeling approach, \citet{brugnach05} provides an initial example of how such process sensitivity analysis might work.
\section{Summary}
In this paper, we described a novel method for representing and developing simulation models of complex ecological and environmental systems.  The representation builds on a two-part ontology that includes entities and processes.  With this formalism, we represent quantitative models in a way that approximates how scientists think about systems.  We also develop reusable libraries of generic entities and processes based on existing ecological knowledge.  We claim that these libraries link model development to existing theory, facilitate model evaluation, and expedite model construction.

Two features contribute to the novelty of our modeling approach.  First, we view model construction as search through a space of possible models.  Second, we instantiate model elements from generic components.  The space of theoretically plausible models and the generic components are defined in libraries of domain specific entities and processes.  We expect this approach to facilitate the construction of ecological models both for theoretical development and for environmental assessment and management.
\section{Acknowledgements}
This research was supported by NSF Grant No. IIS-0326059 and by NTT Communication Science Laboratories, Nippon Telegraph and Telephone Corporation.

\end{document}